\documentclass[11pt,oneside,fleqn]{article}

\usepackage[ansinew]{inputenc}
\usepackage[mathscr]{eucal}
\usepackage{amsmath,amssymb,amsthm}
\usepackage{graphicx}
\usepackage{cite}
\usepackage{hyperref}

\allowdisplaybreaks

\hypersetup{colorlinks, linkcolor=blue, citecolor=blue, urlcolor=blue}
\makeatletter
\long\def\@makecaption#1#2{%
  \vskip\abovecaptionskip\footnotesize
  \sbox\@tempboxa{#1. #2}%
  \ifdim \wd\@tempboxa >\hsize
    #1. #2\par
  \else
    \global \@minipagefalse
    \hb@xt@\hsize{\hfil\box\@tempboxa\hfil}%
  \fi
  \vskip\belowcaptionskip}
\makeatother

\newcommand{\p}{\partial}
\newcommand{\const}{\mathop{\rm const}\nolimits}

\newcommand{\todo}[1][\null]{\ensuremath{\clubsuit}}

\newcommand{\noprint}[1]{}

{\theoremstyle{definition}
\newtheorem{definition}{Definition}

\newtheorem*{remark*}{Remark}
}

\newcommand{\checked}[1][\null]{\ensuremath{\boldsymbol{\surd}}}

\newcommand{\ve}{\varepsilon}

\begin{document}

\par\noindent {\LARGE\bf
Invariant meshless discretization schemes
\par}

{\vspace{4mm}\par\noindent {Alexander Bihlo$^\dag\hspace{0.1mm}^\ddag$
} \par\vspace{2mm}\par}

{\vspace{2mm}\par\noindent {\it
$^{\dag}$~Centre de recherches math\'{e}matiques, Universit\'{e} de Montr\'{e}al, C.P.\ 6128, succ.\ Centre-ville,\\
$\phantom{^\dag}$~Montr\'{e}al (QC) H3C 3J7, Canada\\[0.5ex]
$^{\ddag}$~Department of Mathematics and Statistics, McGill University, 805 Sherbrooke W.,\\
$\phantom{^\ddag}$~Montr\'{e}al (QC) H3A 2K6, Canada
}}

{\vspace{2mm}\par\noindent {\it
$\phantom{^\dag}$~\textup{E-mail}:
bihlo@crm.umontreal.ca
}\par}

\vspace{4mm}\par\noindent\hspace*{8mm}\parbox{140mm}{\small
A method is introduced for the construction of meshless discretization schemes which preserve Lie symmetries of the differential equations that these schemes approximate. The method exploits the fact that equivariant moving frames provide a canonical way of associating invariant functions to non-invariant functions. An invariant meshless approximation of a nonlinear diffusion equation is constructed. Comparative numerical tests with a non-invariant meshless scheme are presented. These tests yield that invariant meshless schemes can lead to substantially improved numerical solutions compared to numerical solutions generated by non-invariant meshless schemes.
}\par\vspace{2mm}

\section{Introduction}

Invariant discretization schemes have received increasing attention over the past 20 years, see e.g.~\cite{bihl12By,bihl12Cy,doro11Ay,doro91Ay,levi06Ay,olve01Ay}. Such schemes are attractive in that they preserve an important property of a system of differential equations, namely its maximal Lie invariance group $G$ or at least a certain physically interesting subgroup of $G$.

One of the most distinct properties of invariant discretization schemes for evolution equations is that they in general require the usage of a moving discretization mesh. The necessity of using non-orthogonal and/or non-stationary meshes considerably complicates the construction and analysis of invariant numerical schemes, especially in the multi-dimensional case. Special techniques can be used to overcome this problem, such as the symmetry-preserving discretization in computational coordinates~\cite{bihl12By,huan10By} or invariant interpolation schemes~\cite{bihl12Cy}. A problem with the former technique is that in the higher-dimensional case quite some computational overhead might be required to construct a proper mapping from the computational domain to the physical space of the system of differential equations.

On the other hand, moving meshes or grids that are adapted to complicated domain geometries which hamper the straightforward use of finite differences or related discretization strategies are not new in the numerical analysis of differential equations. In fact, constructing, storing and modifying discretization grids is costly and hence is attributed as one of the main drawbacks of the otherwise popular finite element method~\cite{ding04Ay,mile12Ay,nguy08Ay}. Indeed, often a large fraction of the computational time spent for the numerical integration of a system of differential equations is consumed by the construction of the discretization mesh itself.

This is why a new class of discretization schemes is steadily growing in importance over the past years, namely so-called \textit{meshless schemes}. The main observations on which meshless methods rest is that no information on the connectivity of the single nodes at which the numerical solution is sought is required in order to discretize a system of differential equations~\cite{ding04Ay,mile12Ay,nguy08Ay}. \emph{All the information needed to approximate the various derivatives is already included in the nodes themselves.} This makes meshless methods attractive to (at least partially) avoid the computational overhead required by the construction of discretization meshes.

It is then obvious to ask whether meshless methods could be employed in the construction of invariant numerical discretization schemes as well. This would allow one to (partially) bypass the complication one faces in attempting to find discrete approximations of a system of differential equations that preserve the symmetries of that system.

It is the purpose of this paper to describe an algorithm for the construction of \emph{invariant meshless discretization schemes}. The key idea on which our construction relies is a property of \emph{equivariant moving frames} to send a given function to an invariant function~\cite{fels98Ay,fels99Ay}. This property was successfully exploited to construct invariant finite difference schemes for partial differential equations starting from non-invariant schemes~\cite{bihl12Cy,kim08Ay,rebe11Ay}. In the present paper we extend this method to meshless discretization schemes.

As what concerns the organization of this paper, in Section~\ref{sec:InvariantMeshlessSchemes} we describe the general procedure for finding invariant meshless discretization schemes. This method is applied in Section~\ref{sec:InvariantDiscretization43DiffusionEquation} to construct an invariant meshless scheme for a nonlinear diffusion equation. Numerical tests comparing the invariant with the non-invariant meshless scheme are carried out in Section~\ref{sec:NumericalTests43DiffusionEquations}. The conclusion and some thoughts for further research directions are presented in Section~\ref{sec:InvariantMeshlessConclusion}.

\section{Invariant meshless discretization schemes}\label{sec:InvariantMeshlessSchemes}

There is not a unique way of constructing meshless approximations to differential equations. In fact, a number of different strategies to discretize a differential equation without or only partial usage of a discretization lattice are used, such as meshless (generalized) finite differences, smooth particle hydrodynamics, the element free Galerkin method or the meshless local Petrov--Galerkin method. For a review of these and further techniques, see e.g.~\cite{bely96Ay,ding04Ay,lisz96Ay,mile12Ay,nguy08Ay} and references therein.

In the present paper we exclusively work with meshless discretizations based on \emph{meshless finite differences}. We stress, though, that similar techniques as introduced below could be applied to other meshless methods that discretize a system of differential equations in the strong form.

Meshless finite differences basically rest on the expansion of a function $u\colon \mathbb{R}^p\to\mathbb{R}$ in a multi-dimensional Taylor series around the node $x^0$,
\[
 u(x)=\sum_{\alpha}\frac{1}{\alpha!}u_{\alpha}|_{x^0}(x-x^0)^\alpha,
\]
where $x=(x_1,\dots,x_p)$ and $\alpha=(\alpha_1,\dots,\alpha_p)$ is a multi-index, $\alpha_j\in\mathbb{N}_0$, $u_\alpha=\partial^{|\alpha|}u/\partial x_1^{\alpha_1}\cdots \partial x_p^{\alpha_p}$, $|\alpha|=\alpha_1+\cdots+\alpha_p$, $\alpha!=\alpha_1!\cdots\alpha_p!$ and $(x-x^0)^\alpha=(x_1-x_1^0)^{\alpha_1}\cdots(x_p-x_p^0)^{\alpha_p}$.

Truncating this series to the $m$th order derivatives one is left with an expansion that includes $s=(p+m)!/(p!m!)$ coefficients. Thus, in theory~$s$ nodes $x^j$ are needed to solve for the~$s$ derivatives $u_{\alpha}$ evaluated at the node $x^0$ from the linear system
\begin{equation}\label{eq:SystemForSolvingForDerivatives}
 u(x^j)=u^j=\sum_{\alpha}\frac{1}{\alpha!}u_{\alpha}|_{x^0}(x^j-x^0)^\alpha,
\end{equation}
$j=1,\dots,s$. The $s$ nodal points $(x^j,u^j)$ are usually chosen to be neighboring points lying inside the $p$-sphere of radius $r$ centered around the node $x^0$. In particular, the nodes are \emph{not} required to lie on a predefined, topologically connected mesh which makes the method truly meshless.

The practical problem that can arise in this construction is that for certain distributions of the $s$ nodes (e.g.\ all points lying on a line), the system~\eqref{eq:SystemForSolvingForDerivatives} cannot be solved for the required derivatives $u_{\alpha}|_{x^0}$. A possible ad hoc remedy for this problem is to include more than $s$ points in the system~\eqref{eq:SystemForSolvingForDerivatives}, i.e.\ to over-determine it~\cite{ding04Ay,lisz96Ay}. The derivatives $u_{\alpha}|_{x^0}$ then follow from the least squares solution of~\eqref{eq:SystemForSolvingForDerivatives}, which reads
\begin{equation}\label{eq:43DiffusionEquationLeastSquareSolution}
 (u^{(m)})^d|_{x^0}=(S^{\rm T}WS)^{-1}S^{\rm T}Wb,
\end{equation}
where
\[
 (u^{(m)})^d|_{x^0}=\left(\begin{array}{c} u^0 \\ u_{x_1}^d \\ \vdots \\ u_{J}^d\end{array}\right),\quad
 S^{\rm T}=\left(\begin{array}{cccc} 1 & 1 & \cdots & 1 \\
  \Delta x_1^1 & \Delta x_1^2 & \cdots & \Delta x_1^k \\
  \vdots & \vdots & \vdots & \vdots \\
  \Delta x_J^1 & \Delta x_J^2 & \cdots & \Delta x_J^k
 \end{array}\right),\quad b=\left(\begin{array}{c} u^1 \\ u^2 \\ \vdots \\ u^k\end{array}\right).
\]
The vector $(u^{(m)})^d|_{x^0}$ contains the $s$ derivatives of the truncated Taylor series evaluated at the node $x^0$ where $u_J^d$ is the highest derivative occurring. The matrix $S$ is build from the associated coefficients of these derivatives where $\Delta x^j_\alpha=(x^j-x^0)^\alpha$. The vector $b$ includes the $k\ge s$ functions values $u^j$ of the nodes inside the radius $r$ of the $p$-sphere centered at $x^0$. A (diagonal) weight matrix $W$ is included in the least squares solution for the geometrical reason to give greater weight to the points $x^j$ closer to $x^0$. More details on this construction can be found in~\cite{ding04Ay,lisz96Ay}. The extension to vector-valued functions $u=(u_1,\dots,u_q)$ is straightforward.

The meshless approximated derivatives $u_\alpha^d$ can be used to discretize a system of differential equations $\mathcal L\colon \Delta_l(x,u^{(m)})=0$, $l=1,\dots,L$, where $u^{(m)}$ includes all the derivatives of $u$ with respect to $x$ up to order $m$ as well as $u$ itself. This leads to a meshless numerical scheme $\mathcal D\colon D_l(x^j,(u^{(m)})^d)=0$, where $(u^{(m)})^d$ denote the discretizations of derivatives in $u^{(m)}$ using Eq.~\eqref{eq:43DiffusionEquationLeastSquareSolution}. If the system $\mathcal L$ includes derivatives of $u$ with respect to the time $t$ then $\mathcal D$ in addition to the meshless spatial derivatives requires a discretizations of the time-dependent derivatives.

\smallskip

We now briefly describe the method of \emph{invariantization} using \emph{moving frames}. An extended discussion can be found in several excellent papers on that subject, including~\cite{fels98Ay,fels99Ay,olve01Ay,olve07Ay,rebe11Ay}.

\begin{definition}\label{def:InvariantDiffusionMovingFrames}
 A (right) \emph{moving frame} $\rho$ is a smooth map $\rho\colon M\to G$ from a manifold $M$ to a finite dimensional Lie group $G$ acting on $M$ with the property that
 \begin{equation}\label{eq:InvariantDiffusionDefinitionMovingFrame}
  \rho(g\cdot z)=\rho(z)g^{-1},
 \end{equation}
for all $z\in M$ and $g\in G$.
\end{definition}
The theorem on moving frame requires a group action to be free and regular in order to guarantee the existence of the moving frame $\rho$. It should be noted that in Definition~\ref{def:InvariantDiffusionMovingFrames} the group $G$ is restricted to be finite dimensional. This restriction is in fact only apparent as the theory of moving frames is already formulated for infinite dimensional Lie (pseudo)groups, see e.g.~\cite{olve08Ay}.

If the group action of $G$ is not free, it can be made free by constructing the moving frame on a jet space ${\rm J}^m={\rm J}^m(M,p)$ of appropriate order $m$. An alternative way of making a group action free is to extend it to the product action on the Cartesian product of copies of the original space~$M$, denoted by $M^\diamond=\{(z^1,\dots,z^k)|\, x^i\ne x^j\: \textup{for all}\: i\ne j\}$ where $z^j=(x^j,u^j)$ are the nodal points. The joint product action is simply the component-wise action, $g\cdot(z^1,\dots,z^k)=(g\cdot z^1,\dots,g\cdot z^k)$, see also~\cite{bihl12Cy,olve01Ay,rebe11Ay}.

Moving frames are determined from a method referred to as \emph{normalization}. For a $r$ dimensional group action, in this procedure one sets up a system of $r$ equations involving the coordinate functions $z$, where $z\in\{x_i,u^{(m)}\}$ in the case $G$ acts on the $m$th order jet space ${\rm J}^m$ or $z\in\{x_i^j,u^j\}$, in case the action of $G$ on the joint product space $M^\diamond$ is considered. The first possibility leads to a moving frame $\rho^{(m)}$ on the jet space ${\rm J}^m$, $\rho^{(m)}\colon {\rm J}^m\to G$ while the second possibility leads to a product frame $\rho^\diamond$ on the space $M^\diamond$, $\rho^\diamond\colon M^\diamond\to G$.

The system of $r$ normalization conditions is chosen in such a manner as to determine a submanifold of ${\rm J}^m$ (or $M^\diamond$) which intersects the group orbits only once and transversally. One usually sets $r$ of the coordinate functions $z$ to appropriately chosen constants, i.e.\ the normalization equations are $z_1=c_1,\dots, z_r=c_r$, although in the discrete case it is beneficial to set combinations of the coordinate functions to constants, see the example in Section~\ref{sec:InvariantDiscretization43DiffusionEquation}. Then one replaces these equations with their respective transformed forms, i.e.\ $Z_1=g\cdot z_1=c_1,\dots, Z_r=g\cdot z_r=c_r$, and solves this algebraic system for the group parameters $g=(\ve_1,\dots,\ve_r)$ in terms of $z$. The result of this construction is the right moving frame $\rho^{(m)}$ (or $\rho^\diamond$).

For the present purpose the most important property of moving frames is that they define a canonical map from a given (non-invariant) function to a $G$-invariant function.

\begin{definition}
 The \emph{invariantization} of a real-valued function $f\colon M\to \mathbb{R}$ using the (right) moving frame $\rho$ is the function $\iota(f)$, which is defined as $\iota(f)(z)=f(\rho(z)\cdot z)$.
\end{definition}


Invariantization is the key for constructing invariant discretization schemes. To accomplish this, the product frame $\rho^\diamond$ is computed using the symmetry group $G$ of the system of differential equations $\mathcal L$ and extending its action to the joint space $M^\diamond$. The product frame should be compatible with the moving frame $\rho^{(m)}$, which is determined for the action of $G$ prolonged to the jet space ${\rm J}^m$. Compatibility means that $\rho^\diamond\to\rho^{(m)}$ in the continuous limit $\{x^j\}\to x$. This is achieved by computing the product frame $\rho^\diamond$ using the discretized form of the normalization conditions that are used to construct the moving frame $\rho^{(m)}$, see~\cite{bihl12Cy,olve01Ay,rebe11Ay} for more details.

With the moving frame $\rho^\diamond$ at hand one can invariantize any standard numerical scheme $\mathcal D$ which approximates a system of differential equations $\mathcal L$. The invariant scheme associated with $\mathcal D$ is
$\iota(\mathcal D)\colon D_l(\iota(x^j),\iota((u^{(m)})^d))=0$, $l=1,\dots,L$, which in the continuous limit $\{x^j\}\to x$ yields the system of differential equations $\mathcal L$ expressed in terms of differential invariants, $\iota(\mathcal L)\colon\Delta_l(\iota(x),\iota(u^{(m)}))=0$. See~\cite{bihl12Cy,kim08Ay,olve01Ay,rebe11Ay} for further details.

The extension of the invariantization procedure to meshless discretization schemes is now straightforward. Once the moving frame $\rho^\diamond$ is determined on the space of nodes $(x^j,u^j)$ it can be applied to the least squares solution~\eqref{eq:43DiffusionEquationLeastSquareSolution} by invariantizing the vectors $(u^{(m)})^d|_{x^0}$ and $b$ and the matrix $S$ in the following way,
\[
 \iota((u^{(m)})^d|_{x^0})=\left(\begin{array}{c} \iota(u^0) \\ \iota(u_{x_1}^d) \\ \vdots \\ \iota(u_{J}^d)\end{array}\right),\quad
 \iota(S)^{\rm T}=\left(\begin{array}{ccc} 1 & \cdots & 1 \\
  \iota(\Delta x_1^1) & \cdots & \iota(\Delta x_1^k) \\
   \vdots & \vdots & \vdots\\
   \iota(\Delta x_J^1) & \cdots & \iota(\Delta x_J^k)
 \end{array}\right),\quad \iota(b)=\left(\begin{array}{c} \iota(u^1) \\ \iota(u^2) \\ \vdots \\ \iota(u^k)\end{array}\right).
\]
The invariantized derivatives in $\iota((u^{(m)})^d|_{x^0})$ and the invariantized nodal points $(\iota(x^j),\iota(u^j))$ are sufficient to invariantize any meshless approximation~$\mathcal D$ of the system~$\mathcal L$. This is done by replacing the occurring nodal points and meshless discrete derivatives in $\mathcal D$ by their respective invariantizations, just in the same way as this is done in the case of finite difference schemes.

An example for this construction is presented in the subsequent section.

\section{Invariant meshless scheme for a nonlinear diffusion equation}\label{sec:InvariantDiscretization43DiffusionEquation}

In this section an invariant meshless scheme is constructed for the nonlinear diffusion equation
\begin{equation}\label{eq:43DiffusionEquation}
 u_t=(u^{-4/3}u_x)_x.
\end{equation}
The meshless Euler forward scheme for Eq.~\eqref{eq:43DiffusionEquation} is given by
\begin{subequations}\label{eq:43DiffusionEquationNonInvariantScheme}
\begin{equation}\label{eq:43DiffusionEquationNonInvariantSchemeEuler}
 \frac{\hat u-u}{\Delta t}= -\frac43u^{-7/3}(u_x^d)^2+u^{-4/3}u_{xx}^d,
\end{equation}
where $\hat u$ and $\check u$ stand for the values of $u=u^0$ at the subsequent and previous time step of the integration, respectively, and $\Delta t=\hat t-t$ is the (constant) time step. The meshless derivatives $u_x^d$ and $u_{xx}^d$ are evaluated at $t$ and $x=x^0$. Likewise, the meshless leapfrog scheme for Eq.~\eqref{eq:43DiffusionEquation} reads
\begin{equation}\label{eq:43DiffusionEquationNonInvariantSchemeLeapfrog}
\frac{\hat u-\check u}{2\Delta t}= -\frac43u^{-7/3}(u_x^d)^2+u^{-4/3}(u_{xx}^d).
\end{equation}
\end{subequations}
In the numerical results reported in Section~\ref{sec:NumericalTests43DiffusionEquations} we use the leapfrog scheme~\eqref{eq:43DiffusionEquationNonInvariantSchemeLeapfrog} and every 20 steps the Euler scheme~\eqref{eq:43DiffusionEquationNonInvariantSchemeEuler} to suppress the computational mode of the leapfrog integration.

In both these schemes the derivatives $u_x^d$ and $u_{xx}^d$ are the meshless approximations of the derivatives $u_x$ and $u_{xx}$. These derivatives are found from the least squares solution~\eqref{eq:43DiffusionEquationLeastSquareSolution}, where in the present one-dimensional case
\[
 (u^{(3)})^d|_{x}=\left(\begin{array}{c} u \\ u_x^d \\ u_{xx}^d \\ u_{xxx}^d\end{array}\right),\quad
 S^{\rm T}=\left(\begin{array}{cccc} 1 & 1 & \cdots & 1 \\
  \Delta x^1 & \Delta x^2 & \cdots & \Delta x^k \\
   \frac12(\Delta x^1)^2 & \frac12(\Delta x^2)^2 & \cdots & \frac12(\Delta x^k)^2 \\
   \frac16(\Delta x^1)^3 & \frac16(\Delta x^2)^3 & \cdots & \frac16(\Delta x^k)^3
 \end{array}\right),\quad b=\left(\begin{array}{c} u^1 \\ u^2 \\ \vdots \\ u^k\end{array}\right),
\]
As a weight matrix we use $W=\mathrm{diag}(\exp(-\mu(\Delta x^j)^2/r^2))$, $j=1,\dots, k$ and $\mu=\const$.

The reason for also including the third derivative in the vector $(u^{(3)})^d|_x$ is that it increases the order of approximation of the derivatives $u_x^d$ and $u_{xx}^d$. If the matrix $S$ is square, it can be directly inferred from the truncated Taylor series that (at least on a uniform grid) including $u_{xxx}^d$ leads to first derivatives with third order accuracy and to second derivatives with second order accuracy. Moreover, in~\cite{ding04Ay} it was shown (again for a uniform grid) that the least square solution ($S$ non-square) invoked for finding $(u^{(m)})^d|_x$ does not degrade the accuracy of the approximation.

We now discuss the invariant meshless approximation of Eq.~\eqref{eq:43DiffusionEquation}.
The diffusion equation~\eqref{eq:43DiffusionEquation} admits a five dimensional maximal Lie invariance algebra $\mathfrak g$ which is generated by
\begin{equation*}
 \p_t,\quad \p_x,\quad 2t\p_t+x\p_x,\quad 2x\p_x-3u\p_u,\quad x^2\p_x-3xu\p_u,
\end{equation*}
see e.g.~\cite{doro03Ay} where an invariant finite difference discretization for~\eqref{eq:43DiffusionEquation} was constructed. The single vector fields in $\mathfrak g$ can be exponentiated to one-parameter Lie groups, which can be composed to yield transformations from the five-parameter maximal Lie invariance group~$G$ of~\eqref{eq:43DiffusionEquation}. The transformations of $G$ acting on $M=\{(t,x,u)\}$ are of the form
\begin{equation}\label{eq:43DiffusionEquationLieGroup}
 T=e^{2\ve_3}(t+\ve_1),\quad X=e^{\ve_3+2\ve_4}\left(\frac{x}{1-\ve_5x}+\ve_2\right),\quad U=e^{-3\ve_4}(1-\ve_5x)^3u.
\end{equation}
The action of $G$ becomes free on the first jet space ${\rm J}^1={\rm J}^1(M,2)$. The moving frame $\rho^{(1)}$ on ${\rm J}^1$ is constructed from the normalization conditions $t=0$, $x=0$, $u=1$, $u_t=1$ and $u_x=0$. This allows one to solve for the group parameters $\ve_1,\dots,\ve_5$. The resulting moving frame $\rho^{(1)}$ is
\begin{align}\label{eq:43DiffusionEquationMovingFrame}
\begin{split}
 &\ve_1=-t,\quad \ve_2=-\frac{x^2u_x+3xu}{3u},\quad \ve_3=\frac12\ln\left(\frac{u_t}{u}\right),\\ &\ve_4=\ln\left(\frac{u_x^{4/3}}{xu_x+3u}\right),\quad \ve_5=\frac{u_x}{xu_x+3u}.
\end{split}
\end{align}

We now turn to the construction of the compatible product frame $\rho^\diamond$. The joint product action of $G$ on $M^\diamond$ follows from evaluating~\eqref{eq:43DiffusionEquationLieGroup} at the single nodes $(t^j,x^j,u^j)$, i.e.\
\begin{equation}\label{eq:43DiffusionEquationProductAction}
  T^j=e^{\ve_3}(t^j+\ve_1),\quad X^j=e^{\ve_3+2\ve_4}\left(\frac{x^j}{1-\ve_5x^j}+\ve_2\right),\quad U^j=e^{-3\ve_4}(1-\ve_5x^j)^3u^j,
\end{equation}
and similar on the subsequent and previous time layers. It is readily verified that the joint product action leaves invariant the condition for the nodes to remain fixed during the integration, which is $\hat x^j-x^j=0$. For this reason, fixed nodes do not break the invariance of Eq.~\eqref{eq:43DiffusionEquation} and we can assume that $\hat x^j=x^j$, which we do for the sake of simplicity. The same is true for the equation $t^{j+1}-t^j=0$, which defines the flatness of the time layers. Hence, $t^{j+1}=t^j=t$.

Before constructing $\rho^\diamond$ it is worthwhile pointing out that the scheme~\eqref{eq:43DiffusionEquationNonInvariantScheme} is already invariant under the action of the one-parameter groups associated with $\ve_1,\dots,\ve_4$. This can be seen as both the variables $t$ and $x^j$ only arise in the differences $\Delta t$ and $\Delta x^j$, which are obviously invariant under translations in $t$ and $x$. At the same time, the scaling properties of the scheme~\eqref{eq:43DiffusionEquationNonInvariantScheme} are the same as of~\eqref{eq:43DiffusionEquation}, which implies the scale invariance of~\eqref{eq:43DiffusionEquationNonInvariantScheme}. Thus, the only symmetry transformation which is violated by~\eqref{eq:43DiffusionEquationNonInvariantScheme} is associated with the group parameter $\ve_5$, i.e.\ inversions in $x$. This is why we construct the invariantization map only for this group parameter.

In the moving frame $\rho^{(1)}$, the component $\ve_5$ follows from the normalization condition $u_x=0$. This normalization condition is replaced by $u_x^d=0$ to guarantee the compatibility of the moving frame $\rho^\diamond$ with $\rho^{(1)}$ in the continuous limit. The problem when using the least square solution~\eqref{eq:43DiffusionEquationLeastSquareSolution} to obtain $u_x^d$ is that it will be very hard (or even impossible) to find the moving frame component $\ve_5$ as the normalization procedure boils down to solving a high order polynomial equation for $\ve_5$. This is why we use a less accurate approximation of $u_x^d$ to compute $\ve_5$. In particular, using the Taylor series expansions $u^r=u+u_x^d(x^r-x)$ and $u^l=u-u_x^d(x-x^l)$ is sufficient to determine $u_x^d$ at $x$, where $(x^r,u^r)$ and $(x^l,u^l)$ are the nodes lying immediately to the left and to the right of $x$. This leads to the usual centered difference approximation
\begin{equation}\label{eq:43DiffusionEquationNormalizationDiscrete}
 u_x^d=\frac{u^r-u^l}{x^r-x^l}.
\end{equation}
Note that on a non-uniform grid it is not guaranteed that this approximation is second order accurate. However, it is numerically verified in the following section that the invariant meshless scheme which follows from the moving frame that employs the approximation~\eqref{eq:43DiffusionEquationNormalizationDiscrete} for $u_x^d$ is more accurate than the associated non-invariant scheme~\eqref{eq:43DiffusionEquationNonInvariantScheme}.

Computing the $\rho^\diamond$-component $\ve_5$ from the normalization~\eqref{eq:43DiffusionEquationNormalizationDiscrete} leads to the cubic equation
\begin{equation}\label{eq:43DiffusionEquationCubicEquation}
 \frac{u^r(x^r)^3-u^l(x^l)^3}{x^r-x^l}\ve_5^3-3\frac{u^r(x^r)^2-u^l(x^l)^2}{x^r-x^l}\ve_5^2+3\frac{u^rx^r-u^lx^l}{x^r-x^l}\ve_5-\frac{u^r-u^l}{x^r-x^l}=0.
\end{equation}
This equation is solved using the explicit formula for the roots which gives three distinct solutions. In the numerical tests we carried out only one of the roots was real, which was then used as the moving frame parameter $\ve_5$.

This apparent non-uniqueness of the component $\ve_5$ in $\rho^\diamond$ already arises when computing $\ve_5$ in the moving frame $\rho^{(1)}$ from the normalization $u_x=0$, which in addition to the solution $\ve_5=u_x/(xu_x+3u)$ also has the solution $\ve_5=1/x$. This solution for $\ve_5$ cannot be used as then the normalizations $x=0$, $u=1$ and $u_t=1$ could not be solved for $\ve_2$, $\ve_3$ and $\ve_4$, respectively. Indeed, it is verified that in the continuous limit $x^r-x\to0$, $x-x^l\to0$, Eq.~\eqref{eq:43DiffusionEquationCubicEquation} has the root $1/x$ of multiplicity 2 and $u_x/(xu_x+3u)$ as a simple root. This shows that the moving frames $\rho^\diamond$ and $\rho^{(1)}$ are indeed compatible.

The invariant meshless counterpart to the scheme~\eqref{eq:43DiffusionEquationNonInvariantScheme} is given by
\begin{subequations}\label{eq:43DiffusionEquationInvariantScheme}
\begin{equation}
 \frac{\iota(\hat u)-\iota(u)}{\Delta t}= -\frac43\iota(u)^{-7/3}\iota(u_x^d)^2+\iota(u)^{-4/3}\iota(u_{xx}^d),
\end{equation}
and
\begin{equation}
 \frac{\iota(\hat u)-\iota(\check u)}{2\Delta t}= -\frac43\iota(u)^{-7/3}\iota(u_x^d)^2+\iota(u)^{-4/3}\iota(u_{xx}^d).
\end{equation}
\end{subequations}
The invariantized discrete derivatives follow from solving Eq.~\eqref{eq:43DiffusionEquationLeastSquareSolution} using the invariantized expressions for $(u^{(3)})^d|_x$, $S$ and $b$, which are, respectively,
\[
 \iota((u^{(3)})^d|_x)=\left(\begin{array}{c} \iota(u) \\ \iota(u_x^d) \\ \iota(u_{xx}^d) \\ \iota(u_{xxx}^d)\end{array}\right),\quad
 \iota(S)^{\rm T}=\left(\begin{array}{ccc} 1 & \cdots & 1 \\ \iota(\Delta x^1) & \cdots & \iota(\Delta x^k) \\ \frac12\iota(\Delta x^1)^2 & \cdots & \frac12\iota(\Delta x^k)^2 \\ \frac16\iota(\Delta x^1)^3 & \cdots & \frac16\iota(\Delta x^k)^3
 \end{array}\right),\quad \iota(b)=\left(\begin{array}{c} \iota(u^1) \\ \iota(u^2) \\ \vdots \\ \iota(u^3)\end{array}\right).
\]
Likewise, the weight matrix $W$ is invariantized to give $\iota(W)=\mathrm{diag}(\exp(-\mu\,\iota(\Delta x^j)^2/r^2))$. In all these formulas, $\ve_5$ is the real solution of Eq.~\eqref{eq:43DiffusionEquationCubicEquation} and we have
\[
 \iota (x^j)=\frac{x^j}{1-\ve_5x^j},\quad \iota (u^j)=(1-\ve_5x^j)^3u^j,\quad \iota (\hat u)=(1-\ve_5x)^3\hat u,\quad \iota (\check u)=(1-\ve_5x)^3\check u.
\]

\section{Numerical tests}\label{sec:NumericalTests43DiffusionEquations}

We compare the invariant scheme~\eqref{eq:43DiffusionEquationInvariantScheme} against the non-invariant meshless scheme~\eqref{eq:43DiffusionEquationNonInvariantScheme} by carrying out numerical tests with the following three exact solutions of Eq.~\eqref{eq:43DiffusionEquation},
\begin{align}\label{eq:ExactSolutions43DiffusionEquation}
\begin{split}
 & u_1=(2c_1x-3c_1^2t+c_2)^{-3/4},\\
 & u_2=\left(\frac{(x+c_1)^2}{t+c_2}+c_3(t+c_2)^2\right)^{-3/4},\\
 & u_3=(c_1x+c_2)^{-3},
\end{split}
\end{align}
where $c_1,c_2,c_3$ are arbitrary constants. The third solution is a stationary solution. For these and further solutions of Eq.~\eqref{eq:43DiffusionEquation}, see~\cite{ibra94Ay,poly04Ay}.

In all the numerical experiments reported we solve Eq.~\eqref{eq:43DiffusionEquation} with the invariant meshless scheme~\eqref{eq:43DiffusionEquationInvariantScheme} and the non-invariant meshless scheme~\eqref{eq:43DiffusionEquationNonInvariantScheme}. We carry out the numerical integration on the interval $L=[1,2]$ and choose the constants $c_1$, $c_2$ and $c_3$ so that the respective exact solution $u_e$ is not singular within $L$. On this interval, we first create an equally-spaced grid with $N=40$ grid points. Each of the grid point is then perturbed by adding a Gaussian distributed random number with zero mean and standard deviation $0.1\cdot\Delta x$, where $\Delta x$ is the spacing of the initial uniform grid. Dirichlet boundary conditions are used with the values of $u^j$ at the boundaries given by the corresponding value of the exact solution $u_1$, $u_2$ or $u_3$.

Ten independent integrations using ten different realizations of the above described grid generation procedure are carried out for $1000$ time steps of the size $\Delta t=0.001$. The time step is rather small so as to avoid numerical instability in the course of the integration. Larger time steps could be used if implicit schemes would be employed instead of explicit schemes in~\eqref{eq:43DiffusionEquationInvariantScheme} and~\eqref{eq:43DiffusionEquationNonInvariantScheme}. The resulting root mean square errors
\[
 \textup{rmse}=\sqrt{\frac{1}{n}\sum_{j=1}^n(u_n^j-u_e^j)^2},
\]
are computed after each integration, where $u_n^j$ and $u_e^j$ are the numerical and exact solutions of Eq.~\eqref{eq:43DiffusionEquation} at $t=1$ at the nodal point $x^j$, respectively. The averaged root mean square errors for the three solutions~\eqref{eq:ExactSolutions43DiffusionEquation} for the non-invariant and the invariant scheme are reported in Table~\ref{tab:NumericalResults43DiffusionEquation} and denoted by $\textup{rmse}_{nis}$ and $\textup{rmse}_{is}$, respectively.
\begin{table}[ht!]
\centering
\caption{Root mean square errors for the exact solutions~\eqref{eq:ExactSolutions43DiffusionEquation} of the nonlinear diffusion equation~\eqref{eq:43DiffusionEquation}.}
\begin{tabular}{l|c|c|c|}
\cline{2-4}
 & $u_1$ & $u_2$ & $u_3$ \\
\hline
\multicolumn{1}{|c|}{$\textup{rmse}_{nis}$} & $3.91\cdot 10^{-4}$ & $2.39\cdot 10^{-3}$ & $2.56\cdot 10^{-3}$\\
\hline
\multicolumn{1}{|c|}{$\textup{rmse}_{is}$} & $5.90\cdot 10^{-5}$ & $1.39\cdot 10^{-4}$ & $0$ \\
\hline
\multicolumn{1}{|c|}{$\textup{rmse}_{is}/\textup{rmse}_{nis}\cdot 100$} & $15.1\%$ & $5.8\%$ & $0\%$\\
\hline
\end{tabular}
\label{tab:NumericalResults43DiffusionEquation}
\end{table}

In the first run (solution $u_1$), $c_1=c_2=0.1$, in the second run (solution $u_2$), $c_1=c_3=0$ and $c_2=10$ and $c_1=c_2=0.1$ in the third run (solution $u_3$). It is worthwhile pointing out that the weight matrices $W$ were chosen differently for the invariant and the non-invariant scheme. The reason for this is that the invariantization of $\Delta x^j$ enters the weight matrix of the invariant scheme. To facilitate the comparison of the results, we have tuned the parameter $\mu$ in each of the runs of the invariant scheme so that the entries in the weight matrices of both the invariant and the non-invariant scheme are of the same order of magnitude.

Table~\ref{tab:NumericalResults43DiffusionEquation} shows that the invariant scheme is able to better approximate the exact solution at $t=1$ in all three test cases. On top of that, for the case of the stationary solution $u_3$ we found that the invariant scheme approximates the exact solution up to machine precision.

\begin{figure}[ht!]
 \centering
 \includegraphics[scale=0.8]{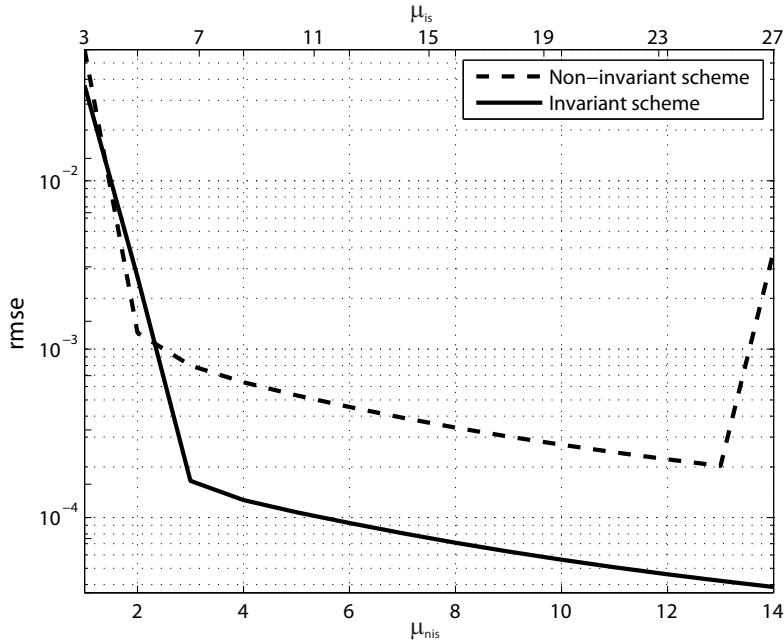}
 \caption{Sensitivity study of the invariant scheme~\eqref{eq:43DiffusionEquationInvariantScheme} and the non-invariant scheme~\eqref{eq:43DiffusionEquationNonInvariantScheme} with respect to the parameter $\mu$ in the weight matrix $W$. Upper $x$-axis: $\mu$ used in the invariant scheme. Lower $x$-axis: $\mu$ used in the non-invariant scheme.}
 \label{fig:diff43SensitivityW}
\end{figure}

As a further sensitivity test we run several integrations of the invariant scheme~\eqref{eq:43DiffusionEquationInvariantScheme} and the non-invariant scheme~\eqref{eq:43DiffusionEquationNonInvariantScheme}, respectively, and vary the parameter $\mu$ in the weight matrices $W$. This parameter controls the influence of distant grid points in the meshless approximation of the discrete derivatives of $u$ at the center node. We use $u_1$ as the exact solution in these runs. In Fig.~\ref{fig:diff43SensitivityW} we depict the result of this sensitivity study. As was discussed before, it is necessary to use different $\mu$ in the invariant and non-invariant integrations due to the different magnitudes of $\Delta x^j$ and $\iota (\Delta x^j)$. It can be seen from Fig.~\ref{fig:diff43SensitivityW} that the invariant numerical scheme~\eqref{eq:43DiffusionEquationInvariantScheme} gives (substantially) better integration results over virtually all values of the parameter $\mu$.

\begin{figure}[ht!]
 \centering
 \includegraphics[scale=0.8]{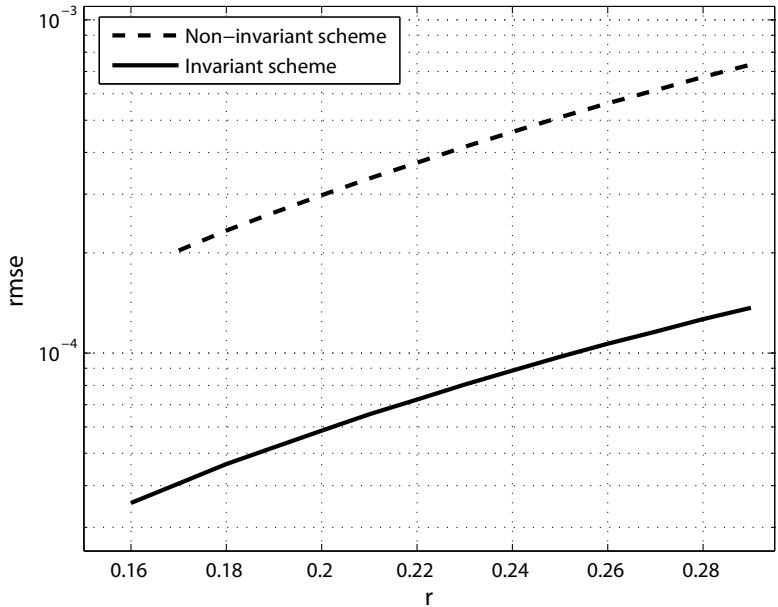}
 \caption{Sensitivity study of the invariant scheme~\eqref{eq:43DiffusionEquationInvariantScheme} and the non-invariant scheme~\eqref{eq:43DiffusionEquationNonInvariantScheme} with respect to the parameter $r$.}
 \label{fig:diff43SensitivityR}
\end{figure}

The final sensitivity test we carry out is with respect to the parameter $r$, i.e.\ the radius within which grid points are used to compute the meshless approximations of the partial derivatives of $u$ at the node $x$. The result of this study using $u_1$ as the exact solution is depicted in Fig.~\ref{fig:diff43SensitivityR}. Again, varying $r$ the invariant scheme is (substantially) better than the non-invariant. Moreover, for the lowest value of $r$ chosen, $r=0.16$, the non-invariant scheme~\eqref{eq:43DiffusionEquationNonInvariantScheme} did not converge, whereas the invariant scheme~\eqref{eq:43DiffusionEquationInvariantScheme} produced the best root mean square error of all the integrations. This means that fewer nodal points are needed to compute a stable approximation of the derivatives for the invariant scheme in the present example.

\section{Conclusion}\label{sec:InvariantMeshlessConclusion}

In this paper we developed a technique for the construction of invariant meshless discretization schemes. The key of this method is the application of the moving frame invariantization map to meshless discrete derivatives which ultimately boils down to the invariantization of a system of truncated Taylor series expansions.

Despite practically demonstrated solely for a one-dimensional evolution equation, the method of invariant meshless discretization as introduced in this paper is particularly suitable for multi-dimensional problems. Differential equations with more than one space dimensions posed a severe problem for the invariant scheme construction machinery available so far due to the necessity of using non-orthogonal, possibly moving discretization meshes. Consequently, most of what is known by today about invariant discretization schemes has been learned from the consideration of ordinary differential equations or single (1+1)-dimensional evolution equations~\cite{doro11Ay,doro03Ay,kim08Ay,levi06Ay,rebe11Ay}. It was only recently that methods for the construction of invariant numerical schemes for multi-dimensional systems of partial differential equations were developed, see e.g.~\cite{bihl12Cy,bihl12By}.

The construction of invariant meshless numerical integrators thus seems to be attractive for several reasons. From the point of view of invariant numerical schemes, it is beneficial to have one more method available that allows one to construct discretization schemes with symmetry properties for equations in any space dimension. In turn, from the side of meshless methods it is interesting to show that ideas from the field of geometric numerical integration can be successfully implemented into such methods. It was shown in this paper that the preservation of qualitative properties of a differential equation in a meshless approximation can substantially increase the quality of the scheme. The invariant discretization we constructed for a nonlinear diffusion equation is able to better reproduce several exact solutions of this equation in practically all the parameter ranges that can be tuned in the scheme. We thereby also demonstrated that preserving symmetries in a numerical integrator is not solely an academic problem.

It will be instructive to apply the proposed technique to multi-dimensional discretization problems and compare invariant meshless schemes against other types of invariant numerical schemes, both in terms of accuracy and computational cost. Also, certain symmetries (e.g.\ Galilean boosts) cannot be preserved on fixed discretization meshes. Moving grid points can lead to strongly distorted meshes and are thus on the risk to deteriorate the quality of the numerical solution or to slow down the convergence rate. This is what generally happens to Lagrangian integration schemes and, as a matter of fact, most invariant numerical schemes preserving Galilean invariance are Lagrangian integrators~\cite{doro11Ay,doro03Ay}. On the other hand, it is known that certain meshless methods are to some degree insensitive regarding the distribution of the nodes. It will therefore be informative to compare invariant meshless methods with discretizations that employ classical or invariant moving meshes such as those constructed in~\cite{bihl12By,huan10By}.

\section*{Acknowledgements}

This research was supported by the Austrian Science Fund (FWF), project J3182--N13. It is a pleasure to thank Professor Roman O.\ Popovych for helpful remarks on the manuscript.

{\footnotesize\setlength{\itemsep}{0ex}

}

\end{document}